\documentclass[letter]{./aa}
\usepackage{natbib}
\usepackage{graphicx}
\begin{document}
   \title{The intensity contrast of solar granulation:\\
          comparing Hinode SP results with MHD simulations}
   \titlerunning{Intensity contrast of solar granulation}
   \author{S. Danilovic \inst{1}\and A. Gandorfer \inst{1} \and
           A. Lagg M.\inst{1} \and Sch{\"u}ssler\inst{1} \and
           S.K. Solanki \inst{1} \and A. V{\"ogler}\inst{2}
           \and Y. Katsukawa \inst{3} \and S. Tsuneta \inst{3}
          }

   \institute{Max-Planck-Institut f\"ur Sonnensystemforschung,
              Max-Planck-Stra{\ss}e 2, 37191 Katlenburg-Lindau, Germany
              \and
              Sterrekundig Instituut, Utrecht University,
              Postbus 80 000, 3508 TA Utrecht, The Netherlands
          \and
              National Astronomical Observatory of Japan,
              2-21-1 Osawa, Mitaka, Tokyo 181-8588, Japan
             }

   \date{\today}

  \abstract
  {The contrast of granulation is an important quantity
  characterizing solar surface convection.}
  {We compare the intensity contrast at 630~nm,
  observed using the Spectro-Polarimeter (SP) aboard the {{\it
  Hinode}} satellite, with the 3D radiative MHD simulations of
  V{\"o}gler \& Sch{\"u}ssler (2007).}
  {A synthetic image from the simulation is degraded using a
  theoretical point-spread function of the optical system,
  and by considering other important effects.}
  {The telescope aperture and the obscuration by the secondary mirror
  and its attachment spider, reduce the simulated contrast from
  14.4\% to 8.5\%. A slight effective defocus of the instrument
  brings the simulated contrast down to 7.5\%, close to the observed
  value of 7.0\%.}
  {A proper consideration of the effects of the optical system and a slight defocus, lead to sufficient degradation of the
   synthetic image from the MHD simulation, such that the contrast reaches almost the observed value. The remaining small discrepancy can be
   ascribed to straylight and slight imperfections of the instrument,
   which are difficult to model. Hence, Hinode SP data are consistent
   with a granulation contrast which is predicted by 3D radiation  MHD
   simulations.}
 \maketitle
%

\section{Introduction}

The root-mean-square of the normalized continuum intensity
fluctuations, within an area on the solar disk, is determined
mainly by the intensity variation between the bright granules and
the darker intergranular lanes, and thus is usually referred to as
the {\it granulation contrast.}  It is a key property of solar
surface convection because it is connected with the temperature
difference between rising (granules) and descending gas masses
(intergranules), and thus related to the efficacy of the
convective energy transport.

Reliable measurements of the granulation contrast are notoriously
difficult, since the observed contrast suffers significantly from
image degradation by the optical system and, most importantly in
the case of ground-based telescopes, by seeing and straylight
effects due to the terrestrial atmosphere. To deduce the `true
granulation contrast' from measurements, we need to deconvolve the
observed images by considering the modulation transfer through the
optical setup (telescope, instruments, detector, etc.), in
addition to through the terrestrial atmosphere. The quantitative
effects of the latter are poorly understood and it is therefore
unsurprising that the reconstructed values of the granulation
contrast, documented in the literature, cover a broad range
\citep[see, e.g., Table 2 in][]{Sanchez-Cuberes:etal:2000}. This
remains true even if additional constraints, such as measurements
of the intensity profile across the lunar limb during a partial
eclipse, are used to derive the effective point-spread function
\citep[e.g.,][]{Levy:1971, Deubner:Mattig:1975, Nordlund:1984,
Sanchez-Cuberes:etal:2000}.

In the case of balloon-borne stratospheric observations, the
influence of the atmosphere is negligible. Even under these
conditions, however, the scatter of the reconstructed contrast
values is large \citep{Bahng:Schwarzschild:1961,
Pravdyuk:etal:1974, Altrock:1976, Edmonds:Hinkle:1977,
Wittmann:Mehltretter:1977, Schmidt:etal:1979, Wittmann:1981},
presumably because the proper consideration of the instrumental
effects is nontrivial \citep{Durrant:etal:1983}.

Although the range of contrast values derived from observations is
quite wide, the values predicted by 3D radiative HD/MHD
simulations are significantly higher
\citep[e.g.][]{Stein:Nordlund:2000, Voegler:etal:2005,
Wedemeyer:2007}. This is unsurprising, since the horizontal
resolution of such simulations (10--40~km) is far better than what
can be achieved observationally at present. To compare the
observed contrast values with predictions from simulations, the
synthetic images derived from simulations must be degraded in the
same way as the observations, taking into account the effects of
the optical system and, in the case of ground-based telescopes,
the terrestrial atmosphere. This procedure is prone to the same
uncertainties as the deconvolution of observations; the result
depends strongly on assumptions about the nature of atmospheric
seeing and straylight \citep{Nordlund:1984, Schuessler:etal:2003,
Rybak:etal:2006} and we are unable to infer if the simulation
predictions are consistent with observations, or otherwise.
However, the good agreement between observed and simulated
spatially averaged spectral line bisectors indicates that the
simulated intensity contrasts are probably close to actual values
\citep{Asplund:etal:2000d}.

A new era has begun with the launch of the 50-cm Solar Optical
Telescope \citep[SOT,][]{Tsuneta:etal:2008} on the {\it Hinode\/}
satellite \citep{Kosugi:etal:2007}.  The good performance and low
straylight level of the spectro-polarimeter
\citep[SP,][]{Lites:etal:2001}and complete absence of atmospheric
effects, enable a far more reliable determination of the
granulation contrast at $630\,$nm, the wavelength at which the SP
operates. In this {\it Letter\/}, we compare the intensity
contrast of a {\it Hinode\/} SP continuum map of the quiet Sun
with predictions of MHD simulation.


\section{Observations and simulation data}

We determine the observed contrast from a map of the continuum
intensity at $630\,$nm wavelength obtained on Jan. 16, 2007
(12:10:10 -- 13:36:49 UT), using the scan mode of the {\it
Hinode\/} SP with an exposure time of $4.8\,$s per slit position.
The map covers a field of $163\arcsec\times 164\arcsec$ of `quiet'
Sun, close to disk center. The slit width, the sampling step size,
and the CCD pixel pitch all correspond to $0.16\arcsec$. The data
were reduced using standard routines that correct for various
instrumental effects \citep{Lites:etal:2007b}.  After reduction,
we find an rms contrast of the continuum intensity of 7.0\%.

The 3D MHD simulation snapshot that we consider represents the
saturated (statistically stationary) state of the dynamo run C of
\citet{Voegler:Schuessler:2007}. The computational box contains
$648\times 648 \times 140$ cells; it corresponds to a physical
size on the Sun of $4.86\times 4.86$~Mm$^2$ in the horizontal and
1.4~Mm in the vertical direction, the latter ranging from about
900~km below to 500~km above continuum optical depth unity at
$630\,$nm $(\tau_{630}=1)$. The simulation has been run with
non-grey radiative transfer using the {\sl MURaM} code
\citep{Voegler:2003, Voegler:etal:2005}. With a horizontal
grid-cell size of 7.5~km, this simulation had one of the highest
resolutions achieved so far. The snapshot that we use, has an
average unsigned vertical magnetic field of about 7~G at
$\tau_{630}=0.1$, which we consider as a reasonable representation
of the `quiet' Sun.  The corresponding original (unsmeared)
continuum image at $630\,$nm shows an rms contrast (standard
deviation divided by mean value) of 14.4\%.  The particular choice
of snapshot was not critical for the results presented here.
Considering a number of other snapshots and simulations of
different spatial resolution or amount of magnetic flux in the
simulation box, we found contrast values that were typically
between 14\% and 15\%.

\section{Modeling of the system PSF}

\begin{figure}
    \centering
    \includegraphics[width=0.9\hsize,angle=90]{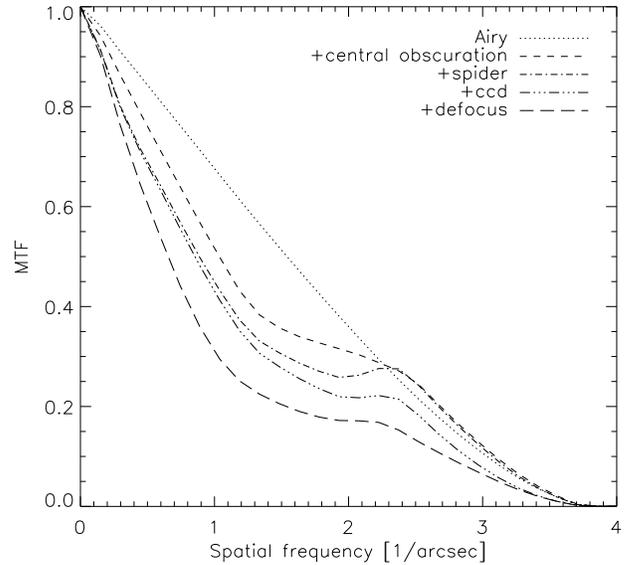}
    \caption{Change of the Modulation Transfer Function when the effects
    of the different parts of the optical system are sequentially taken
    into account: telescope aperture (`Airy'), central obscuration,
    secondary mirror and spider, sampling (CCD), and a
    defocus of $1.5\,$mm (about 9 steps of the focus mechanism). }
    \label{fig:mtfs}
\end{figure}

The difference between the observed and original contrast of the
simulation amounts to more than a factor of two (7.0\% vs.
14.4\%). Since there is no atmospheric distortion of the image, we
investigate whether instrumental effects alone can account for
this significant degradation.

The intrinsic resolution of the simulated images considered here
(equivalent to $0.01\arcsec$ pixel size) is considerably better
than the resolution of the 50-cm {\it Hinode\/} SOT; the
degradation of the synthetic image based on the simulation result
can therefore be modeled by applying an effective point-spread
function (PSF).  We convolved our simulated intensity maps in the
2D spatial domain with different PSFs, which represented a more
and more realistic optical systems.  To this end, we modeled the
telescope and the transfer optical path down to the spectrograph
entrance slit using the commercial optical design software
ZEMAX\footnote{www.zemax.com}. While the telescope was modeled
with the nominal SOT surface parameters, the transfer
(refocussing) path was modeled in paraxial approximation, which,
however, has no influence on the final results of our
calculations. ZEMAX calculates the PSF as the Fourier transform of
the wavefront in the exit pupil of the system.

We describe the contrast degradation by the optical system using
the Modulation Transfer Function (MTF), which is the Fourier
transform of the point-spread function. In this way, the various
effects can be considered step by step, as indicated below.
Figure~\ref{fig:mtfs} shows the corresponding change of the system
MTF.

{\sl Telecope aperture:} The {\it Hinode\/} SOT is an $f/9.1$ Gregory
system with an aperture of $50\,$cm, which corresponds to a cut-off at
$1/0.26\arcsec$. Higher spatial wave numbers are not transmitted by the
system, while low to intermediate wave numbers are transmitted with
relatively high contrast.

{\sl Central obscuration and spider:} Important contributors to
the contrast reduction are the central obstruction of the
telescope by the secondary mirror, which produces a linear
obscuration of $0.344=17.2\,{\rm cm}/50\,{\rm cm}$, and the
diffraction by the three spider elements, which are of $4\,$cm
width each, holding the secondary. The effect of the central
obstruction is that the highest spatial wave numbers are largely
unaffected, while the intermediate wave numbers are significantly
damped. The spider structure adds to this damping.

{\sl Spatial sampling:} Since the data considered here are
obtained using the spectro-polarimeter, we have to consider the
sampling and integration effects of the slit and the CCD detector.
The sampling and the integration effects of the detector are
approximated by powers of sinc functions \citep{Boreman:2001}.

{\sl Defocus:} Low-order optical aberrations damp the MTF at low
spatial wave numbers. Since the {\it Hinode} Focal Plane Package
(FPP) is a complex system, it is out of the scope of this {\sl
Letter} to model in detail all relevant contributions from
optical, optomechanical, and electronic parts of the system. We
therefore consider only the effect of a defocus in the plane of
the SP slit. It turns out that a small amount of defocus can
reduce significantly the intensity contrast, while leaving the
fine-scale resolution of the system unaffected.  The focus of the
{\it Hinode\/} science instruments is controlled by means of a
common reimaging lens that can be shifted in a range of $\pm
25\,$mm, with a step of $0.17\,$mm.

In addition, we investigated the effect of the (lossy) JPEG
compression which was applied to the images. The change of the rms
contrast turned out to be negligible, even for high compression
factors.

\section{Degradation of the simulation data}

To compare with the {\it Hinode\/} result, we convolved the
intensity maps calculated from the simulation with the different
PSFs resulting from modeling the {\it Hinode\/} SOT as described
in the previous section.  After convolution, the degraded images
were binned to represent the sampling and integration effects of
the finite slit width, step size of the spatial sampling, and the
pixel size.

\begin{figure}
    \centering
    \includegraphics[width=0.8\hsize,angle=90]{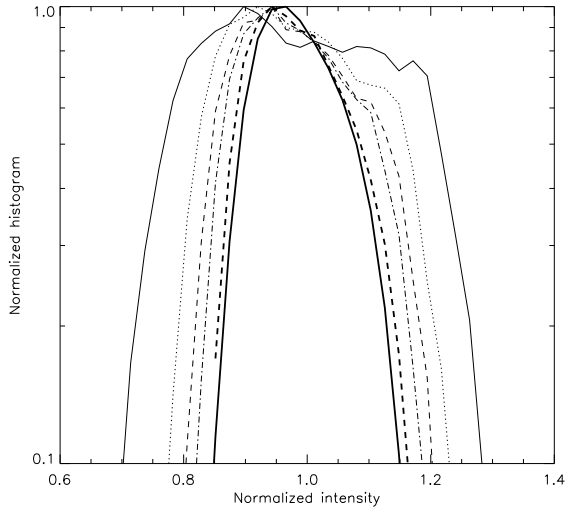}
    \caption{Normalized histograms of the continuum-intensity values at
    630~nm, obtained for 35 equal bins between 0.6 and 1.4 of the mean
    intensity.  Shown are results for the original synthetic image from
    the simulation snapshot (outermost, solid curve) and for the
    degraded images corresponding to the MTFs shown in
    Fig.~\ref{fig:mtfs} (same line styles as used there). The two thick
    inner curves represent the degraded synthetic image assuming
    a defocus of 1.5~mm (thick dashed curve) and the  continuum map
    observed with {\it Hinode\/}~SP (thick solid curve).}
    \label{fig:conthist}
\end{figure}

Figure~\ref{fig:conthist} shows the evolution of the histogram of
intensity values in the simulation image as the various effects of
the optical system are introduced. The rms contrast values for the
various steps are listed in Table~\ref{tab:rms}. The basic
telescope effects (primary aperture, obscuration, and spider)
reduce the rms contrast to 8.7\%, presumably since the small-scale
intensity structure in the intergranular lanes is lost. There is
almost no further contrast degradation due to sampling and
rebinning (CCD), which indicates that the remaining contrast is
determined mainly by spatial scales that are significantly larger
than the critical sampling limit of the diffraction pattern. Once
the fine detail in the granular lanes is no longer resolved, the
contrast is dominated by the intensity differences between the
granules and the average intensity of the intergranular lanes,
which is almost unaffected by the integrating effect of the
detector pixels. Similar values of the contrast reduction up to
this step were obtained by \citet{Orozco:etal:2007b}.

\begin{table}
\caption{Values of the rms intensity contrast after applying the PSFs
  corresponding to the MTFs shown in Fig.~\ref{fig:mtfs}.}
\label{tab:rms}
\centering
\begin{tabular}{l c}
\hline \hline
effects taken into account & rms [\%] \\
\hline
none & 14.4 \\
Airy & 10.9 \\
+ central obscuration & 9.6 \\
+ spider & 8.7 \\
+ CCD & 8.5 \\
+ defocus (1.5$\,$mm) & 7.5 \\
\hline
\end{tabular}
\end{table}

\ifnum 2<1
\begin{table}
\caption{Contrast as a function of telescope defocus.}
\label{tab:defocus}
\centering
\begin{tabular}{c c }
\hline \hline
defocus value [mm] & rms [\%] \\ 
\hline 
0.10 & 7.3 \\
0.15 & 7.1 \\
0.16 & 7.0 \\
0.18 & 6.6 \\
\hline 
\end{tabular}
\end{table}
\fi

\begin{figure}
    \centering
    \includegraphics[width=0.8\hsize,angle=90]{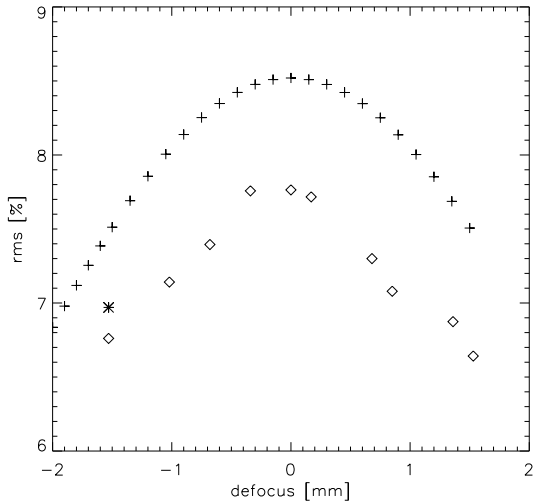}
    \caption{Intensity contrast as a function of telescope defocus.
Shown are results for a SP data set used taken on Jan. 16, 2007,
10:54:14--11:06:52 UT (diamonds) and the contrast values for the
degraded simulation image (crosses). The value for the large map
used to compare with the simulation is indicated by the asterisk,
assuming that the location of the optimal focus was the same as
for the other dataset. This also gives an indication of the
observational uncertainty.}
\label{fig:focus_pos}
\end{figure}

The difference between the rms contrast degraded so far (8.5\%)
and the observed contrast (7.0\%) is still significant.  Another
factor that contributes to contrast degradation is a slight
defocus of the SP, which affects mainly the low to intermediate
spatial wave numbers.  We have no direct information about the
amount of defocus for the dataset considered here; we, however,
obtained empirical evidence about the effect of defocus on the
intensity contrast using a number of SP datasets for different
focus positions, taken only an hour earlier than the map analyzed
here.  Apart from the exposure time of $1.6\,$s per slit position
and the smaller size of the maps, all other observational
parameters are the same.  For every focus position, a map that
covers a field of $3.2\arcsec\times 82\arcsec$ was generated.  The
corresponding contrast values given in Fig.~\ref{fig:focus_pos}
confirm the rather sensitive dependence on focus position. The
figure also shows the dependence of the contrast of the simulation
image on the defocus value. The large observational map with a
contrast of 7.0\% (indicated by the asterisk) was taken with a
defocus of 8-9 steps (1.36$\,$mm -- 1.53$\,$mm) from the optimal
position in Fig.~\ref{fig:focus_pos}. The degraded simulation
image corresponding to about this defocus value (1.5$\,$mm) has a
contrast of 7.5\%.

\begin{figure*}
    \centering
    \includegraphics[width=0.95\hsize,angle=0]{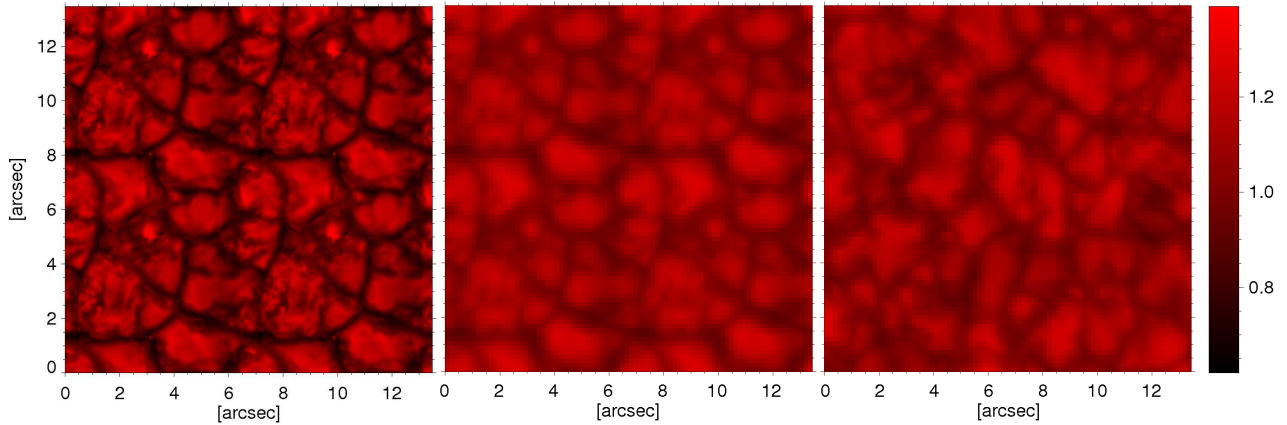}
    \caption{Continuum images at $630\,$nm from the simulation snapshot
    with original resolution {\it (left)} and after degrading {\it
    (middle)}, in comparison with a detail of the observed
    Hinode/SP map of the same size {\it (right)}. The periodic
    simulation box is shown fourfold for better visibility.}
    \label{fig:cont}
\end{figure*}

As an illustration of the image degradation, Fig.~\ref{fig:cont} shows
the original continuum image from the simulation (left panel, the
periodic simulation box is drawn fourfold for the sake of better
visibility) and the final degraded image (middle).  A subset of
the observational map of the same size is shown in the right panel,
using the same gray scale.

The remaining deviation of 0.5\% from the observed value of about
7\%, and the general shift between the curves for simulation and
observation in Fig.~\ref{fig:focus_pos}, can be ascribed to
various degrading factors that are not included in our analysis.
These include the effects of straylight, of all other low-order
optical aberrations apart from defocus, and of the read-out
electronics (such as transfer efficiency and pixel-to-pixel
crosstalk), which reduce the contrast, but are difficult to model
reliably.  To illustrate the extreme sensitivity of the
granulation contrast to low-order aberrations and straylight we
calculated, as an example, in ZEMAX the effect of astigmatism and
coma, which were equally distributed in such a way that the total
system shows an rms wave-front error of 0.044 wavelengths. This
corresponds to the number given by Suematsu et al. (2008) and
represents a system with a remarkable Strehl ratio of 0.93. This
small effect is nevertheless sufficient to reduce further the
contrast from 8.5\% to 8.1\% in optimal focus position. If no
further degrading effects are present, a straylight level of only
4.7\% would be sufficient to bring the contrast down to the
observed value of 7.8\% (for optimal focus), and to 7.0\% for the
presumptive defocus value of the dataset studied here,
respectively.

\section{Conclusion}

We find that consideration of the basic optical properties of the
{\it Hinode} SOT/SP system and a slight defocus are sufficient to
bring the degraded contrast of a 3D radiative MHD simulation and
the observed rms continuum contrast at $630\,$nm, almost into
agreement. The remaining discrepancy can be ascribed to the
combined, minor effects of straylight, other low-order optical
aberrations, and instrument electronics. Hence, Hinode SP data are
consistent with a granulation contrast at 630~nm of 14-15\%, as
predicted by the simulations.

\begin{acknowledgements}
Hinode is a Japanese mission developed and launched by ISAS/JAXA, with
NAOJ as domestic partner and NASA and STFC (UK) as international
partners. It is operated by these agencies in co-operation with ESA and
NSC (Norway).
\end{acknowledgements}


\bibliographystyle{aa}

\end{document}